# ANALYSIS OF CYBER-ATTACK CLASS-SPECIFIC TOPOLOGICAL METRICS IN NATURAL VISIBILITY GRAPH REPRESENTATIONS OF NETWORK TRAFFIC

**Ph.D., Ali Melih KANCA**
Department of Computer Engineering, Faculty of Computing and Informatics Sciences, Karabuk University, Karabuk, Türkiye
ORCID ID: https://orcid.org/0000-0003-3271-7463
melihkanca@karabuk.edu.tr

**Prof. İlker TÜRKER**
Computer Engineering, Faculty of Computing and Informatics Sciences, Karabuk University, Karabuk, Türkiye
ORCID ID: https://orcid.org/0000-0001-7577-4658
iturker@karabuk.edu.tr

**ABSTRACT**

Natural Visibility Graph (NVG)-based representations provide a promising approach for capturing structural patterns in sequential network traffic. However, whether different cyber-attack classes exhibit distinctive topological signatures in such representations remains insufficiently understood. This study investigates the discriminative and structural characteristics of NVG-based network traffic representations using the CSE-CIC-IDS2018 dataset. Seventy-six numerical traffic features were independently transformed into NVGs within overlapping frames of 40 observations, and ten graph-theoretic metrics were extracted from each graph, resulting in 760 topological descriptors per frame. The discriminative capability of these representations was evaluated using a multi-branch convolutional neural network (CNN) with stratified five-fold cross-validation. The model achieved an average accuracy of 96.20% and a Matthews correlation coefficient (MCC) of 0.9566. To characterize class-specific topological differences, Kruskal–Wallis and Mann–Whitney U tests were combined with Benjamini–Hochberg false discovery rate correction and effect-size measures. Of the 10,640 attack-versus-benign comparisons, 7,777 (73.1%) remained statistically significant after FDR correction, with 4,844 exhibiting large Cliff's delta effects. The strongest global differences were predominantly associated with backward-traffic and packet-length-related features combined with connectivity, clustering, and centrality measures. These findings indicate that NVG-derived representations can provide strong discriminative capability while revealing class-dependent topological patterns associated with different cyber-attack classes.

## INTRODUCTION

The increasing diversity and complexity of cyber-attacks have made network-based intrusion detection an important component of modern cybersecurity systems. Network traffic generated by benign activities and malicious behaviors such as denial-of-service (DoS), distributed denial-of-service (DDoS), brute-force, and botnet attacks exhibits different temporal and statistical characteristics. Conventional intrusion detection approaches generally represent these characteristics through raw flow features or feature vectors derived directly from network traffic. Although such representations can provide effective classification performance, they may not explicitly capture the structural relationships that emerge among sequential traffic observations, which may limit the characterization of attack-specific behavioral patterns.

Graph-based representations provide an alternative way to capture structural information from sequential network traffic. In particular, the Natural Visibility Graph (NVG) transforms a time series into a graph by connecting observations according to their visibility relationships (Lacasa et al., 2008) . This transformation preserves structural relationships within the original sequence while enabling the use of graph-theoretic measures to characterize its topological properties (Iacovacci & Lacasa, 2016) .

Consequently, NVG-based representations provide an alternative perspective for analyzing network traffic beyond the statistical values of individual flow features.

Although graph-based approaches have increasingly been investigated for cybersecurity and intrusion detection (Yao et al., 2019; Zhong et al., 2024; Zhu & Lu, 2022; Zola et al., 2022) , existing studies have primarily focused on improving detection or classification performance. Comparatively less attention has been given to understanding the structural characteristics underlying graph-based traffic representations. In particular, for NVG-based representations, it remains insufficiently understood whether different cyber-attack classes exhibit distinctive topological signatures and which combinations of traffic features and graph-theoretic measures are most strongly associated with these class-specific structures. Addressing this gap is important for moving beyond classification performance toward a more interpretable characterization of attack-related traffic behavior.

To address this gap, this study develops a systematic feature-wise NVG framework for representing network traffic. Individual numerical traffic features are independently transformed into Natural Visibility Graphs, and a set of graph-theoretic metrics is extracted from each resulting graph. The resulting topological descriptors are evaluated through two complementary stages. First, a convolutional neural network (CNN) is employed to determine whether the NVG-derived representations can effectively discriminate among different traffic classes. Second, non-parametric statistical analyses and effect-size measures are used to identify traffic-feature–topology combinations that exhibit distinctive differences between attack and benign traffic.

Using 76 numerical traffic features and 10 graph-theoretic metrics, the proposed framework generates a comprehensive topological representation of network traffic. The classification capability of these representations is evaluated using stratified five-fold cross-validation. The resulting CNN achieves an average accuracy of 96.20% and a Matthews correlation coefficient (MCC) of 0.9566, demonstrating the strong discriminative capability of the NVG-derived representations. Furthermore, attack-specific statistical analysis based on Kruskal–Wallis testing, Mann–Whitney U testing, Cliff's delta, and false discovery rate (FDR) correction reveals substantial differences across many traffic-feature–topology combinations, providing a basis for identifying distinctive topological signatures of individual attack classes.

## MATERIALS AND METHODS

### Dataset

The experiments were conducted using the CSE-CIC-IDS2018 dataset, which contains network traffic generated by benign activities and multiple cyber-attack scenarios (*IDS 2018 | Datasets | Research | Canadian Institute for Cybersecurity | UNB*, n.d.). The complete combined dataset was considered rather than using a manually selected subset of traffic features. After preprocessing, 76 numerical traffic features were retained for the proposed analysis.

To preserve the temporal organization of the traffic observations, the data were divided into overlapping frames of 40 observations with a step size of 20, corresponding to a 50% overlap between consecutive frames. This process resulted in 4,999 frames. Each frame was subsequently processed independently for NVG construction and topological feature extraction.

Each frame was assigned a single class label based on the ground-truth labels of the observations contained within the frame. A majority-voting strategy was applied, whereby the most frequently occurring class label was assigned to the frame. When benign and attack labels occurred with equal frequency, the non-benign label was prioritized to preserve the presence of malicious traffic within the frame.

### Overall Analysis Framework

The proposed framework consists of four main stages: (i) traffic framing, (ii) feature-wise NVG construction, (iii) topological feature extraction, and (iv) classification and statistical characterization. For each frame, the 76 numerical traffic features were independently represented as time series and transformed into Natural Visibility Graphs. Ten graph-theoretic metrics were then calculated from each NVG, resulting in 760 topological descriptors per frame. These descriptors were subsequently evaluated using a CNN-based multi-class classifier and complementary non-parametric statistical analyses.

The classification stage was used to evaluate the discriminative capability of the NVG-based representations, whereas the statistical stage focused on identifying traffic-feature–topology combinations associated with attack-specific structural differences.

**Natural Visibility Graph Construction**

For each numerical traffic feature, a Natural Visibility Graph (NVG) was constructed independently within each frame. Given a time series $x_1, x_2, \dots, x_N$, each observation was represented as a graph node. Two observations $i$and $j$, where $i < j$, were connected when the straight line joining $x_i$and $x_j$remained above all intermediate observations. Formally, an edge exists between nodes $i$and $j$ if, for every intermediate observation:

$$x_k < x_i + \frac{x_j - x_i}{j - i}(k - i),\ i < k < j.$$

This procedure transforms the temporal evolution of each traffic feature into a graph structure while preserving visibility relationships among observations.

**Topological Feature Extraction**

Ten graph-theoretic metrics were extracted from each NVG to characterize different aspects of its topology: Average Degree, Density, Average Clustering Coefficient, Average Shortest Path Length, Diameter, Mean Betweenness Centrality, Mean Closeness Centrality, Mean Eigenvector Centrality, Modularity, and Number of Communities.

These metrics collectively characterize complementary aspects of graph structure, including connectivity, local organization, global path characteristics, node centrality, and community organization. Since 76 traffic features were independently transformed into NVGs, the resulting representation contains 76 × 10 = 760 topological descriptors per frame.

**CNN-Based Classification**

A multi-branch one-dimensional convolutional neural network (CNN) was employed to evaluate the discriminative capability of the NVG-derived representations. Each traffic feature was assigned to an independent CNN branch whose input consisted of its corresponding 10 topological metrics. Each branch included a one-dimensional convolutional layer with 32 filters and a kernel size of 3, followed by batch normalization, dropout, and flattening. The outputs of all branches were concatenated and passed to a fully connected layer with 128 neurons before the final softmax classification layer.

The model was trained using the Adam optimizer with a learning rate of 0.001, sparse categorical cross-entropy loss, a batch size of 32, and 20 training epochs. Model performance was evaluated using stratified five-fold cross-validation. Accuracy, precision, recall, F1-score, and Matthews correlation coefficient (MCC) were used to assess classification performance.

**Statistical Analysis**

Statistical analysis was performed to characterize the differences in topological descriptors across traffic classes. First, the Kruskal–Wallis test (Chicco et al., 2025) was applied independently to each of the 760 traffic-feature–topology combinations to identify descriptors exhibiting significant differences among classes. Benjamini–Hochberg false discovery rate (FDR) correction was subsequently applied to account for multiple comparisons (Benjamini & Hochberg, 1995). The magnitude of the observed differences was quantified using epsilon-squared ($\epsilon^2$) as an effect-size measure. To further characterize attack-specific signatures, each attack class was compared with benign traffic using the Mann–Whitney U test(Chicco et al., 2025). Benign traffic was used as the common reference group to characterize how the NVG-derived topological structure associated with each attack class deviates from normal network traffic. Cliff's delta (δ), a non-parametric effect-size measure suitable for quantifying group differences without relying on normality assumptions(Meissel & Yao, 2024), was calculated to characterize the magnitude and direction of the differences between attack and benign traffic. Positive values indicate higher descriptor values for attack traffic, whereas negative values indicate lower values. FDR correction was applied to the attack-specific comparisons, and the resulting effect sizes were used to identify traffic-feature–topology combinations with the strongest attack-specific signatures. The magnitude of Cliff's delta was categorized according to the absolute effect size as negligible ($|\delta| < 0.147$), small ($0.147 \leq |\delta| < 0.330$), medium ($0.330 \leq |\delta| < 0.474$), and large ($|\delta| \geq 0.474$).

## RESULTS AND DISCUSSION

### CNN Classification Performance

The discriminative capability of the NVG-derived topological representations was first evaluated using the proposed multi-branch CNN under stratified five-fold cross-validation. The model achieved an average accuracy of 96.20% ± 0.92%, with a weighted precision of 96.37% ± 0.82%, weighted recall of 96.20% ± 0.92%, and weighted F1-score of 96.16% ± 0.89%. The corresponding macro-averaged precision, recall, and F1-score were 91.97% ± 4.42%, 90.26% ± 4.63%, and 90.48% ± 4.42%, respectively. The model also achieved an MCC of 0.9566 ± 0.0105, indicating strong agreement between the predicted and true class assignments. The overall cross-validation performance of the proposed multi-branch CNN is illustrated in Figure 1. The overall cross-validation performance metrics are summarized in Table 1.

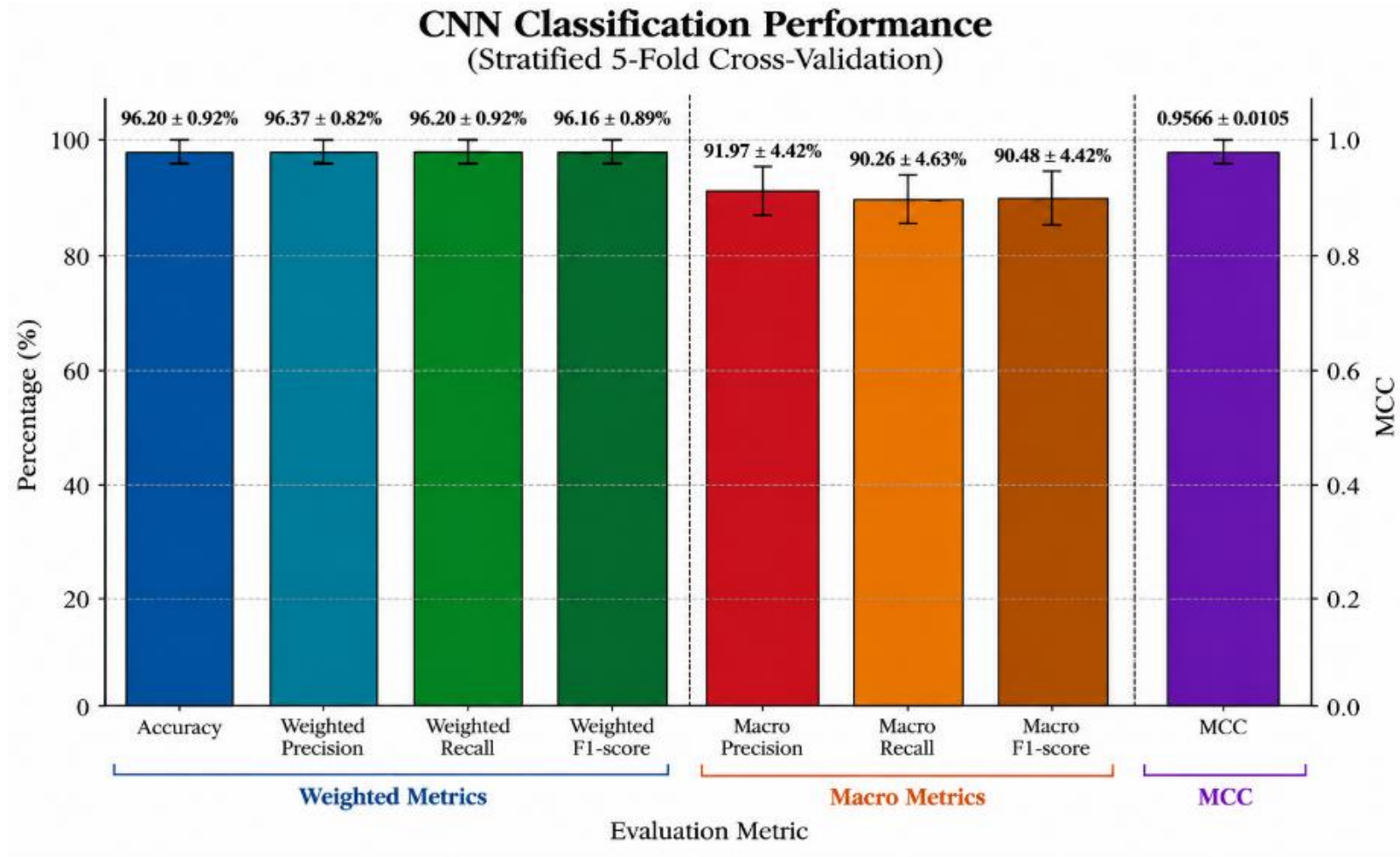

**Figure 1.** CNN classification performance under stratified five-fold cross-validation.

**Table 1.** CNN Cross-Validation Performance.

| Metric | Mean ± Std. |
|---|---|
| Accuracy | 96.20 ± 0.92% |
| Weighted Precision | 96.37 ± 0.82% |
| Weighted Recall | 96.20 ± 0.92% |
| Weighted F1-score | 96.16 ± 0.89% |
| Macro Precision | 91.97 ± 4.42% |
| Macro Recall | 90.26 ± 4.63% |
| Macro F1-score | 90.48 ± 4.42% |
| MCC | 0.9566 ± 0.0105 |

The relatively high weighted and macro-averaged performance indicates that the topological representations preserve substantial class-discriminative information. The difference between the weighted and macro-averaged scores, however, suggests that performance is less uniform across classes, particularly for classes with relatively limited representation. This observation is further examined through the class-level results and confusion matrix presented in the following subsection

**Class-Level Classification Performance**

The pooled out-of-fold predictions provide a more detailed view of class-level performance. Most classes achieved F1-scores above 0.94, with several classes reaching approximately 0.996 or higher. The class-level classification results based on pooled out-of-fold predictions are presented in Table 2.

The corresponding class-level F1-scores are illustrated in Figure 2. In contrast, lower F1-scores were observed for Brute Force-Web, SQL Injection, and particularly DoS attacks-Slowloris. DoS attacks-Slowloris achieved an F1-score of 0.6000, with a recall of only 0.4286; however, this class contains only seven out-of-fold samples. Therefore, its relatively low performance should be interpreted cautiously, as the limited sample size substantially restricts the reliability of the class-level performance estimates.

Overall, the class-level results support the strong aggregate performance observed in the cross-validation experiments while also indicating that classification performance is not uniform across all classes.

**Table 2.** Class-level performance based on pooled out-of-fold predictions**.**

| Traffic Class | Precision | Recall | F1-score | Support |
|---|---|---|---|---|
| Benign | 0.9637 | 0.9746 | 0.9691 | 1,143 |
| Infilteration | 0.9960 | 0.9960 | 0.9960 | 496 |
| Bot | 0.7778 | 0.8750 | 0.8235 | 32 |
| DoS attacks-SlowHTTPTest | 0.8889 | 0.7273 | 0.8000 | 11 |
| DDoS attacks-LOIC-HTTP | 0.9960 | 0.9919 | 0.9939 | 248 |
| DoS attacks-GoldenEye | 0.9140 | 0.9770 | 0.9444 | 87 |
| FTP-BruteForce | 0.9960 | 0.9960 | 0.9960 | 496 |
| DDOS attack-HOIC | 0.9960 | 0.9960 | 0.9960 | 248 |
| DDOS attack-LOIC-UDP | 0.9960 | 0.9960 | 0.9960 | 248 |
| Brute Force -Web | 0.7888 | 0.7984 | 0.7936 | 248 |
| Brute Force -XSS | 1.0000 | 0.9919 | 0.9960 | 248 |
| SQL Injection | 0.8017 | 0.7823 | 0.7918 | 248 |
| DoS attacks-Hulk | 0.9766 | 0.9667 | 0.9716 | 991 |
| DoS attacks-Slowloris | 1.0000 | 0.4286 | 0.6000 | **7** |
| SSH-Bruteforce | 1.0000 | 0.9960 | 0.9980 | 248 |

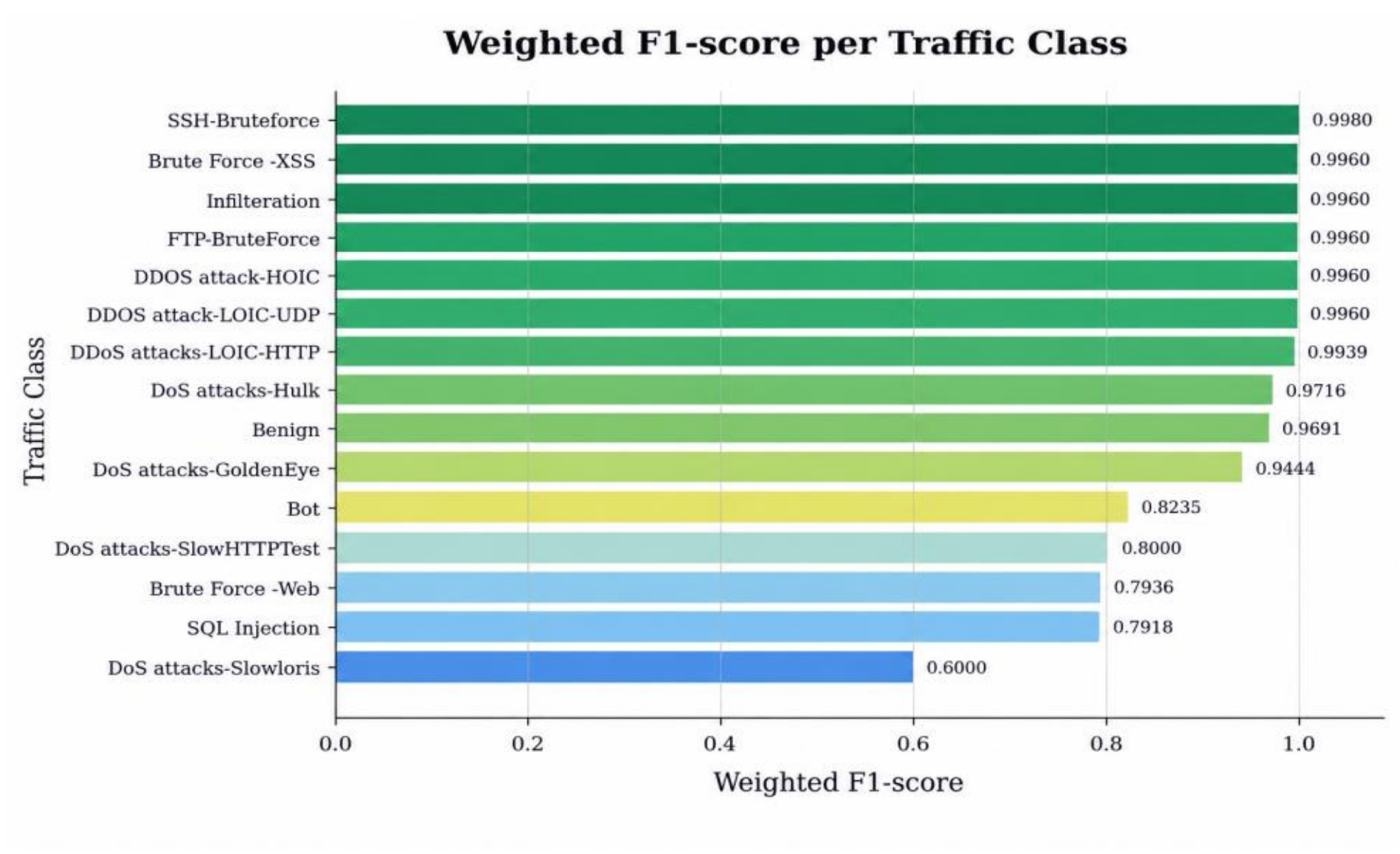

**Figure 2.** Class-level F1-scores based on pooled out-of-fold predictions.

### Global Topological Differences

To investigate whether the extracted topological descriptors differed systematically among traffic classes, a Kruskal–Wallis test was performed for each traffic-feature–topology combination. Among the 760 evaluated traffic-feature–topology combinations, 680 yielded valid Kruskal–Wallis test statistics, of which 677 remained statistically significant after Benjamini–Hochberg FDR correction. The remaining 80 combinations corresponded to eight traffic features across all ten graph metrics and did not yield valid Kruskal–Wallis p-values. This indicates that a large proportion of the extracted topological descriptors exhibit statistically significant differences across traffic classes. These findings indicate that the topological structure induced by individual traffic features varies substantially across traffic classes. In particular, the prominence of backward-traffic and packet-length-related features suggests that the temporal organization of packet exchange is reflected in the resulting NVG topology. The top 15 traffic-feature–topological metric combinations ranked by epsilon-squared effect size are presented in Table 3.

**Table 3.** Top 15 traffic-feature–topological metric combinations ranked by epsilon-squared effect size.

| Rank | Traffic Feature | Graph Metric | Kruskal–Wallis H | $\varepsilon^2$ | FDR |
|---|---|---|---|---|---|
| 1 | TotLen Bwd Pkts | Average Degree | 4153.7653 | 0.8306 | $< 0.001$ |
| 2 | TotLen Bwd Pkts | Density | 4153.7653 | 0.8306 | $< 0.001$ |
| 3 | Subflow Bwd Byts | Average Degree | 4153.7653 | 0.8306 | $< 0.001$ |
| 4 | Subflow Bwd Byts | Density | 4153.7653 | 0.8306 | $< 0.001$ |
| 5 | TotLen Bwd Pkts | Average Clustering Coefficient | 4145.3228 | 0.8289 | $< 0.001$ |
| 6 | Subflow Bwd Byts | Average Clustering Coefficient | 4145.3228 | 0.8289 | $< 0.001$ |
| 7 | Bwd Pkt Len Max | Average Degree | 4081.6991 | 0.8162 | $< 0.001$ |
| 8 | Bwd Pkt Len Max | Density | 4081.6991 | 0.8162 | $< 0.001$ |
| 9 | TotLen Bwd Pkts | Eigenvector Centrality | 4025.7351 | 0.8049 | $< 0.001$ |
| 10 | Subflow Bwd Byts | Eigenvector Centrality | 4025.7351 | 0.8049 | $< 0.001$ |

| 11 | Bwd Pkt Len Max | Average Clustering Coefficient | 4018.1646 | 0.8034 | < 0.001 |
|---|---|---|---|---|---|
| 12 | Bwd Seg Size Avg | Average Degree | 4007.5428 | 0.8013 | < 0.001 |
| 13 | Bwd Seg Size Avg | Density | 4007.5428 | 0.8013 | < 0.001 |
| 14 | Bwd Pkt Len Mean | Average Degree | 4007.5428 | 0.8013 | < 0.001 |
| 15 | Bwd Pkt Len Mean | Density | 4007.5428 | 0.8013 | < 0.001 |

The identical statistics observed for Average Degree and Density arise from their deterministic relationship when graph size is fixed, as all NVGs in the present analysis contain the same number of nodes.

Effect-size analysis further demonstrated that these differences were not limited to statistical significance. The strongest observed effects were associated predominantly with backward-traffic and packet-length-related features. The highest effect size was obtained for TotLen Bwd Pkts × Average Degree and Subflow Bwd Byts × Average Degree, both yielding an epsilon-squared value of 0.8306. These were followed by combinations involving Average Clustering, Density, and Eigenvector Centrality. Overall, the strongest global effects were consistently associated with backward-traffic characteristics, indicating that the temporal organization of reverse-direction traffic is strongly reflected in the resulting NVG topology.

**Attack-Specific Topological Signatures**

To further characterize attack-specific structural signatures, each of the 14 attack classes was independently compared with benign traffic across all 76 traffic features and 10 topological metrics. This resulted in 10,640 attack-versus-benign comparisons, of which 7,777 (73.1%) remained significant after FDR correction.

Among the significant comparisons, 4,844 exhibited large effect sizes, while 1,027 and 1,312 comparisons showed medium and small effects, respectively. Only 594 significant comparisons were categorized as having negligible effects. Thus, approximately 62.3% of the FDR-significant comparisons exhibited large effects, indicating that many of the statistically significant differences were also substantial in magnitude.

Cliff's delta additionally revealed the direction of these differences. Strong positive and negative effect sizes were observed for several traffic-feature–topology combinations, indicating that attack traffic could exhibit either elevated or reduced topological characteristics relative to benign traffic depending on the underlying traffic feature and graph metric. For example, several combinations produced absolute Cliff's delta values greater than 0.99, demonstrating highly pronounced attack-versus-benign differences.

These findings provide empirical evidence that the evaluated attack classes are associated with distinct patterns in the NVG-derived topological descriptors. Importantly, the observed signatures are not restricted to a single graph metric or traffic feature, but emerge from specific interactions between the temporal characteristics of traffic features and the topology of their corresponding visibility graphs.

**Table 4**. Strongest FDR-significant topological signature for each cyber-attack class relative to benign traffic, based on the maximum absolute Cliff's delta (|δ|).

| Attack Class | Strongest Traffic Feature | Graph Metric | Cliff's δ |
|---|---|---|---|
| Infilteration | Fwd Pkts/s | Average Degree | −0.9910 |
| Bot | Fwd Act Data Pkts | Average Clustering | +0.8474 |
| DoS attacks-SlowHTTPTest | Flow IAT Mean | Average Degree | −0.8814 |
| DDoS attacks-LOIC-HTTP | Init Bwd Win Byts | Eigenvector Centrality | −0.9687 |
| DoS attacks-GoldenEye | Fwd IAT Tot | Average Degree | −0.9950 |
| FTP-BruteForce | Pkt Size Avg | Eigenvector Centrality | +0.9946 |
| DDOS attack-HOIC | Bwd Pkt Len Std | Eigenvector Centrality | −0.9721 |
| DDOS attack-LOIC-UDP | Init Bwd Win Byts | Eigenvector Centrality | −0.9284 |
| Brute Force -Web | Flow IAT Std | Average Degree | −0.9994 |
| Brute Force -XSS | Bwd Pkt Len Mean | Average Clustering | −0.9177 |
| SQL Injection | Fwd IAT Tot | Average Degree | −0.9999 |
| DoS attacks-Hulk | ECE Flag Cnt | Average Clustering | +0.7071 |
| DoS attacks-Slowloris | — | — | — |
| SSH-Bruteforce | Tot Bwd Pkts | Eigenvector Centrality | −0.9936 |

As shown in Table 4, the strongest attack-specific topological signatures varied considerably across attack classes in terms of both traffic feature and graph metric. Several attack classes exhibited very large effect sizes, including SQL Injection (Fwd IAT Tot × Average Degree, $\delta = -0.9999$), Brute Force-Web (Flow IAT Std × Average Degree, $\delta = -0.9994$), and FTP-BruteForce (Pkt Size Avg × Eigenvector Centrality, $\delta = 0.9946$). The direction of the effects also differed across attacks, indicating that attack-specific deviations from benign traffic may involve either increases or decreases in the corresponding NVG-derived topological characteristics. No FDR-significant attack-versus-benign comparison was identified for DoS attacks-Slowloris.

## DISCUSSION

The experimental results demonstrate that Natural Visibility Graph representations can provide a meaningful topological characterization of network traffic. The CNN achieved an average accuracy of 96.20% and an MCC of 0.9566, indicating that the temporal structures encoded by the NVGs contain sufficient information to distinguish the evaluated traffic classes. This result supports the first research question and suggests that graph topology can complement conventional flow-level representations in network traffic classification.

The statistical analysis provides further insight into the source of this discriminative capability. The Kruskal–Wallis analysis revealed significant class-dependent differences across a large proportion of the evaluated traffic-feature–topology combinations. The relatively high epsilon-squared values observed for several combinations indicate that these differences are not merely a consequence of large sample sizes, but can also correspond to substantial differences in the distributions of topological descriptors.

The attack-versus-benign analysis provides a more specific interpretation of these findings. More than seventy percent of the evaluated comparisons remained significant after FDR correction, and a substantial proportion of these comparisons exhibited large Cliff's delta effects. This indicates that attack-related differences are reflected in multiple dimensions of graph topology rather than being restricted to a single structural property. The observed positive and negative effect directions further suggest that attack traffic may alter different aspects of the visibility structure depending on the underlying traffic feature.

The frequent appearance of backward-traffic, packet-length, and inter-arrival-related features among the strongest associations is particularly noteworthy. These features describe aspects of packet exchange intensity, directionality, size, and temporal organization. When transformed into visibility graphs, changes in these characteristics can modify connectivity, clustering, path structure, and centrality patterns. Therefore, the results suggest that attack-specific behavior can manifest as changes in the topology induced by the temporal evolution of individual traffic features.

Nevertheless, the class-level results also reveal an important limitation. Classification performance is not uniform across all classes, and classes with very limited numbers of frames exhibit less reliable performance estimates. In particular, DoS attacks-Slowloris contains only seven samples, making its recall and F1-score highly sensitive to individual predictions. Therefore, the corresponding attack-specific statistical findings should be interpreted with respect to class sample size and statistical power.

Overall, the findings indicate that NVG-based representations can serve not only as inputs for high-performance classification but also as a basis for investigating the structural characteristics of different traffic classes. This dual role distinguishes the proposed framework from approaches that evaluate graph representations primarily through their final classification accuracy.

## LIMITATIONS

Several limitations should be acknowledged. First, the analysis was conducted on a single benchmark dataset, and therefore the generalizability of the observed topological signatures to other network environments requires further investigation. Second, some attack classes contain substantially fewer samples than others, limiting the statistical power and reliability of class-specific estimates.

## CONCLUSION

This study investigated whether Natural Visibility Graph representations of network traffic can provide discriminative and interpretable topological signatures for different cyber-attack classes. A feature-wise NVG framework was developed by transforming 76 numerical traffic features into visibility graphs and extracting 10 graph-theoretic metrics from each representation. The resulting 760-dimensional topological representation was evaluated using a multi-branch CNN and complementary statistical analyses.

The classification results demonstrated strong discriminative capability, with an average accuracy of 96.20% and an MCC of 0.9566 under stratified five-fold cross-validation. Beyond classification performance, the statistical analysis revealed substantial class-dependent differences across the extracted topological descriptors. In particular, the attack-versus-benign analysis identified 7,777 FDR-significant comparisons out of 10,640, with 4,844 exhibiting large effect sizes.

These findings indicate that cyber-attack behavior can be reflected in the topology induced by the temporal evolution of individual network traffic features. The results also show that attack-specific signatures are distributed across different combinations of traffic features and graph-theoretic properties rather than being captured by a single topological descriptor. Therefore, NVG-based representations offer a promising framework for combining attack classification with structural characterization of network traffic.

Future work will focus on evaluating the proposed framework across additional datasets and network environments and developing more interpretable methods for identifying the most informative traffic-feature–topology relationships. Further investigation of the relationship between topological signatures and specific attack behaviors may also improve the explainability of graph-based intrusion detection systems.